\documentstyle[prl,twocolumn,aps]{revtex}
\begin{document}
\draft
\author{K.B. Efetov}
\title{Directed Quantum Chaos}
\address{{\it Max-Planck Institut f\"ur Physik komplexer Systeme, Heisenbergstr.1,
70569 Stuttgart, Germany and }\\{\it L.D. Landau Institute for Theoretical
Physics, Moscow, Russia}}
\date{\today }
\maketitle

\begin{abstract}
Quantum disordered problems with a direction (imaginary vector-potential)
are discussed and mapped onto a supermatrix $\sigma $-model. It is argued
that the $0D$ version of the $\sigma $-model may describe a broad class of
phenomena that can be called directed quantum chaos. It is demonstrated by
explicit calculations that these problems are equivalent to problems of
theory of random asymmetric or non-Hermitian matrices. A joint probability
of complex eigenvalues is obtained. The fraction of states with real
eigenvalues proves to be always finite for time reversal invariant systems.
\end{abstract}

New very interesting phenomena may occur in systems with non-Hermitian
quantum Hamiltonians. Although the Hamiltonian of a closed system in
equilibrium must be Hermitian, non-Hermitian models can describe
nonequilibrium processes \cite{fogedby}, open systems connected to
reservoirs \cite{verb}, dynamics of neural networks \cite{somp}, and have
many other applications.

In a recent remarkable work \cite{hatano} Hatano and Nelson considered a
model of particles described by a random Schr\"odinger equation with an
imaginary vector-potential. This model arises as a result of mapping of flux
lines in a $(d+1)$-dimensional superconductor to the world lines of $d$%
-dimensional bosons. Columnar defects introduced experimentally in order to
pin the flux lines \cite{civale} lead to the random potential in the boson
system, whereas the component of the magnetic field perpendicular to the
defects results in the constant imaginary vector-potential $i{\bf h}$ \cite
{nelson}. The prediction made in Ref. \cite{hatano} is that already a
one-dimensional chain of the bosons has to undergo a
localization-delocalization transition as a function of disorder. This
effect is new and very unusual for physics of disordered systems, which
clearly shows that random models with the imaginary vector potentials
deserve a further discussion.

The Hamiltonian $H$ of the simplest model of non-interacting particles with
a disorder and the imaginary vector potential $i{\bf h}$ has the form 
\begin{equation}
\label{a1}H=\left( {\bf \hat p+}i{\bf h}\right) ^2/\left( 2m\right) +U\left( 
{\bf r}\right) 
\end{equation}
where ${\bf \hat p=-}i\nabla $ and $U\left( {\bf r}\right) $ is a random
potential.

Although the imaginary vector-potential $i{\bf h}$ appears quite naturally
in the model of the flux lines, the existence of such a quantity in the
conventional quantum mechanics seems to be rather exotic because it violates
the hermiticity. However, one can imagine situations when this term may be
finite.

Of course, this is not possible for a closed system of the particles. But we
know that taking into account degrees of freedom of an environment can lead
to the non-hermiticity. One of the well known examples is a quantum dot
connected to leads. Instead of considering the whole system one can
eliminate the leads (see, e.g. \cite{zirn}). This results in complex
eigenenergies in the dot; the imaginary parts of the eigenenergies describe
the probability of an escape into the leads.

At the same time, the non-Hermitian Hamiltonian, Eq. (\ref{a1}), differs
from the Hamiltonians for open dots. It is real and contains the vector $%
{\bf h}$ that introduces a direction. Average physical quantities have to
depend on the direction of ${\bf h}$ but the time-reversal symmetry is not
broken. To understand better the physical meaning of the vector ${\bf h}$ it
is instructive to write a lattice Hamiltonian $H_L$ corresponding to $H$ 
$$
H_L=-\frac t2\sum_{{\bf r}}\sum_{\nu =1}^d\left( e^{{\bf he}_\nu }c_{{\bf r+e%
}_\nu ^{}}^{+}c_{{\bf r}}+e^{-{\bf he}_\nu }c_{{\bf r}}^{+}c_{{\bf r+e}_\nu
}\right) 
$$
\begin{equation}
\label{a2}+\sum_{{\bf r}}U\left( {\bf r}\right) c_{{\bf r}}^{+}c_{{\bf r}} 
\end{equation}
where $c^{+}$ and $c$ are creation and annihilation operators, and $\left\{ 
{\bf e}_\nu \right\} $ are the unit lattice vectors. Considering the one
particle Hamiltonian we do not need to specify statistics of the particles.
(Eq. (\ref{a2}) was used in Ref.\cite{hatano} for numerical calculations).

In Eq. (\ref{a2}), the hopping probability along ${\bf h}$ is higher than in
the opposite direction. In other words, the Hamiltonian $H_L$ describes a 
{\it directed} hopping in a random potential. This model can be considered
as a quantum counterpart of a directed percolation model introduced by
Obukhov \cite{obukhov}. Such models may describe properties of disordered
systems without a center of inversion. The finite ${\bf h}$ in Eq. (\ref{a2}%
) can be due to, e.g. a possibility of a virtual tunneling into a dielectric
in an electric field. This would not lead to an imaginary contribution to
the chemical potential as in open quantum dots but might make hopping in
different directions nonequivalent. It is interesting to note that Green
functions used in Ref. \cite{obukhov} contained as the energy spectrum the
combinations $\left( {\bf p+}i{\bf a}\right) ^2$ with a constant vector $%
{\bf a}$, which corresponds to the continuum version of the Hamiltonian, Eq.
(\ref{a1}).

``Conventional'' (non-directed) disordered systems exhibit a variety of
different phenomena. Depending on the geometry of the sample and strength of
disorder one can have conduction, localization or, e.g. quantum chaos. One
can guess that the models described by Eq. (\ref{a1},\ref{a2}) also contain
these effects although the effects may be peculiar. In analogy with the
directed percolation it is reasonable to call the corresponding phenomena 
{\it directed quantum chaos}, {\it directed localization}, etc.

The main results of Ref. \cite{hatano} were derived for $1D$ and $2D$
samples from numerical computations, although some qualitative conclusions
were made considering a one-impurity model. At the same time, very well
developed analytical methods of study of conventional disordered systems
exist and it is highly desirable to develop analogous schemes for the
directed disorder problems. This would help to understand the
localization-delocalization transitions but also consider new phenomena like
directed quantum chaos.

In this Letter it is shown that problems of directed disorder described by
Eqs. (\ref{a1},\ref{a2}) can be reduced to calculations within a non-linear
supermatrix $\sigma $-model. This approach proved to be very useful for a
broad variety of different problems (for a review see\cite{efetov}). The $%
\sigma $-model derived below is applicable in any dimension and differs from
previous ones by a new ${\bf h}$-dependent term. Problems of quantum chaos
correspond to the zero-dimensional version and are most simple for
calculations. Some new results are obtained for this case classified below
as {\it directed quantum chaos}.

Recently a ``ballistic'' $\sigma $-model was derived by averaging over
either rare impurities or energy \cite{muz}. Apparently, proper ${\bf h}$%
-terms can be written for this case, too.

Although the imaginary vector potential ${\bf h}$ enters Eqs. (\ref{a1},\ref
{a2}) almost in the same way as the physical vector-potential ${\bf A}$, the
presence of $i$ changes considerably the derivation of the $\sigma $-model.
A simple replacement $\left( e/c\right) {\bf A}\rightarrow i{\bf h}$ in the $%
\sigma $-model of Refs. \cite{efetov,efetov1}, would lead to the absence of
the ground state. This is because eigenenergies of the Hamiltonians, Eq. (%
\ref{a1}) or Eq. (\ref{a2}), are not necessarily real and retarded $%
G_\epsilon ^R$ and $G_\epsilon ^A$ Green functions can have now poles
everywhere in the complex plane of $\epsilon $. As a result, these functions
cannot be written in the usual form of convergent Gaussian integrals, which
was a necessary step in derivation of the $\sigma $-model.

First of all one should choose a proper quantity to calculate. According to
a discussion of Ref. \cite{hatano} an important information can be obtained
studying a joint probability $P\left( \epsilon ,y\right) $ of real $\epsilon
_k^{\prime }$ and imaginary $\epsilon _k^{^{\prime \prime }}$ parts of
eigenenergies or, in other words, the density of complex eigenenergies. This
function is introduced as 
\begin{equation}
\label{a3}P\left( \epsilon ,y\right) =V^{-1}\left\langle \sum_k\delta \left(
\epsilon -\epsilon _k^{\prime }\right) \delta \left( y-\epsilon _k^{\prime
\prime }\right) \right\rangle 
\end{equation}
where $V$ is the volume of the system, the sum is taken over all
eigenstates, and $\left\langle ...\right\rangle $ stands for averaging over
the disorder. In contrast to the average density of states in conventional
disordered systems the function $P\left( \epsilon ,y\right) $ at ${\bf h\neq 
}0$ distinguishes between localized and extended states. For the localized
states it is proportional to $\delta \left( y\right) $ but is a non-trivial
function for extended ones.

It is convenient to rewrite Eq. (\ref{a3}) as 
\begin{equation}
\label{a4}P\left( \epsilon ,y\right) =\frac 1{\pi V}\lim _{\gamma
\rightarrow 0}\left\langle \sum_k\frac{\gamma ^2}{\left[ \left( \epsilon
-\epsilon _k^{\prime }\right) ^2+\left( y-\epsilon _k^{\prime \prime
}\right) ^2+\gamma ^2\right] ^2}\right\rangle 
\end{equation}
which allows to express the function $P\left( \epsilon ,y\right) $ in terms
of a Gaussian integral over supervectors $\psi _k$. This can be done using
the fact that the ratio in the sum in Eq. (\ref{a4}) can be composed of the
elements of the matrix $M_k^{-1}$, where 
$$
M_k=\left( 
\begin{array}{cc}
i\gamma -\left( \epsilon -\epsilon _k\right) & -i\left( y-\epsilon
_k^{\prime \prime }\right) \\ 
i\left( y-\epsilon _k^{\prime \prime }\right) & i\gamma +\left( \epsilon
-\epsilon _k^{\prime }\right) 
\end{array}
\right) 
$$
In order to replace the integrals over all $\psi _k$ by integrals over
supervector fields $\psi \left( {\bf r}\right) $ one should use vectors $%
u_k\left( {\bf r}\right) $ and $v_k\left( {\bf r}\right) $ and their
conjugates $\bar u_k\left( {\bf r}\right) $ and $\bar v_k\left( {\bf r}%
\right) $%
$$
u_k=\frac 12\left( 
\begin{array}{c}
\phi _k\left( 
{\bf r}\right) +\bar \phi _k^{*}\left( {\bf r}\right) \\ \phi _k\left( {\bf r%
}\right) -\bar \phi _k^{*}\left( {\bf r}\right) 
\end{array}
\right) ,_{}v_k=\frac 12\left( 
\begin{array}{c}
\phi _k\left( 
{\bf r}\right) -\bar \phi _k^{*}\left( {\bf r}\right) \\ \phi _k\left( {\bf r%
}\right) +\bar \phi _k^{*}\left( {\bf r}\right) 
\end{array}
\right) 
$$
where $\phi _k\left( {\bf r}\right) $ and $\bar \phi _k\left( {\bf r}\right) 
$ are right and left eigenfunctions of the Hamiltonian $H$, Eq. (\ref{a1})
(or $H_L$, Eq. (\ref{a2})), the symbol $^{*}$ is complex conjugation. With
the vectors $u_k\left( {\bf r}\right) $ and $v_k\left( {\bf r}\right) $ one
can make Fourier expansion for any $2$-component vector field.

Using $8$-component supervectors $\psi \left( {\bf r}\right) $ with exactly
the same structure as those in Refs.\cite{efetov1,efetov} one can rewrite
Eq.(\ref{a4}) in terms of a functional integral with the Lagrangian ${\cal L}
$%
\begin{equation}
\label{a5}{\cal L}=-i\int \bar \psi \left( {\bf r}\right) {\cal H}\psi
\left( {\bf r}\right) d{\bf r} 
\end{equation}
where the $8\times 8$ matrix operator ${\cal H}$ has the form 
\begin{equation}
\label{a6}{\cal H}=H^{\prime }-\epsilon +i\gamma \Lambda +H^{\prime \prime
}\Lambda _1+iy\Lambda _1\tau _3 
\end{equation}
In Eq. (\ref{a6}), $H^{\prime }=\frac 12\left( H+H^{+}\right) $ and $%
H^{\prime \prime }=\frac 12\left( H-H^{+}\right) $ are Hermitian and
anti-Hermitian parts of the Hamiltonian, respectively. In the continuum
version $H^{\prime \prime }={\bf h\nabla }/m$. The matrices $\Lambda $ and $%
\tau _3$ are the same as in Refs. \cite{efetov1,efetov} and the matrix $%
\Lambda _1$ anticommutes with $\Lambda $ being equal to 
\begin{equation}
\label{a7}\;\Lambda _1=\left( 
\begin{array}{cc}
0 & {\bf 1} \\ {\bf 1} & 0 
\end{array}
\right) 
\end{equation}
where ${\bf 1}$ is the $4\times 4$ unit matrix. A proper pre-exponential
term is rather lengthy and is not written here.

Further steps of derivation of the $\sigma $-model are standard. One
averages over the random potential and decouples the effective interaction
by integration over $8\times 8$ supermatrices $Q$. The integral over the
eigenvalues of $Q$ is calculated using the saddle-point approximation. The $%
\sigma $-model is finally obtained by expansion over ${\bf \nabla }Q$ and $%
{\bf h}$. As a result, the function $P\left( \epsilon ,y\right) $ takes the
form 
\begin{equation}
\label{a8}P\left( \epsilon ,y\right) =-\lim _{\gamma \rightarrow 0}\frac{\pi
\nu ^2}{4V}\int A\left[ Q\right] \exp \left( -F\left[ Q\right] \right) DQ 
\end{equation}
where the free energy functional $F\left[ Q\right] $ is written as%
$$
F\left[ Q\right] =\frac{\pi \nu }8\int STr[D_0\left( \nabla Q+{\bf h}\left[
Q,\Lambda _1\right] \right) ^2 
$$
\begin{equation}
\label{a9}-4Q\left( \gamma \Lambda +y\Lambda _1\tau _3\right) ]d{\bf r} 
\end{equation}
In Eq. (\ref{a9}), $D_0$ is the classical diffusion coefficient, $\left[
.,.\right] $ is the commutator, $STr$ is the supertrace, and $\nu $ is the
density of states of the system without disorder at ${\bf h=}0$. The
pre-exponential functional $A\left[ Q\right] $ equals 
$$
A\left[ Q\right] =\int [\left( Q_{42}^{11}\left( {\bf r}\right)
+Q_{42}^{22}\left( {\bf r}\right) \right) \left( Q_{24}^{11}\left( {\bf r}%
^{\prime }\right) +Q_{24}^{22}\left( {\bf r}^{\prime }\right) \right) - 
$$
\begin{equation}
\label{a10}-\left( Q_{42}^{21}\left( {\bf r}\right) +Q_{42}^{12}\left( {\bf r%
}\right) \right) \left( Q_{24}^{21}\left( {\bf r}^{\prime }\right)
+Q_{24}^{12}\left( {\bf r}^{\prime }\right) \right) ]d{\bf r}d{\bf r}%
^{\prime } 
\end{equation}
Numeration of matrix elements in Eq. (\ref{a10}) is the same as in Refs. 
\cite{efetov1,efetov}; Eqs. (\ref{a8}-\ref{a10}) can be used in any
dimension. The supermatrix $Q$ has the same symmetry as $Q$ for the
orthogonal ensemble in Refs. \cite{efetov1,efetov}. The free energy
functional $F\left[ Q\right] $, Eq. (\ref{a9}), has two additional terms
with respect to the functional used for ``conventional'' disorder problems.
These terms contain the matrix $\Lambda _1$, which leads to new effective
``external fields'' in the free energy. We see that the replacement of the
physical vector-potential ${\bf A}$ by the imaginary quantity $i{\bf h}$
changes the symmetry of $F\left[ Q\right] $. This reflects the fact that $%
{\bf A}$ and $i{\bf h}$ violate different physical symmetries.

Of course, one can include in Eqs. (\ref{a1},\ref{a2}) the vector-potential $%
{\bf A}$, which would result in a standard term in Eq. (\ref{a9}) describing
a magnetic field. If this field is strong enough one can use Eqs. (\ref{a8}-%
\ref{a10}) as before but in this case $Q$ is a supermatrix with the
structure corresponding to the unitary ensemble of Refs. \cite
{efetov1,efetov}. This is a system with the broken time reversal invariance.

Calculation of the function $P\left( \epsilon ,y\right) $ can be carried out
using methods presented in Ref. \cite{efetov}. One can try to use
renormalization group methods for study of $2D$ case (corresponding to $3D$
case for flux lines), transfer-matrix method for $1D$ case (thick films with
parallel line defects and magnetic field) or calculate definite integrals
over $Q$ for the $0D$ (flux lines in long cylinders). Leaving $1D$ and $2D$
cases for future study let us consider now the $0D$ case classified here as
directed quantum chaos.

In this limit one should integrate over $Q$ assuming that this variable does
not depend on coordinates. Then, the gradient terms in $F\left[ Q\right] $
in Eq. (\ref{a9}) can be omitted and integration over ${\bf r}$, ${\bf r}%
^{\prime }$ in Eq. (\ref{a10}) easily performed. The $0D$ form of $F\left[
Q\right] $ obtained in this way allows to make a very interesting conclusion
even without starting explicit computation. This can be done comparing Eq. (%
\ref{a9}) with results of a recent work \cite{fyodorov} in which the
supersymmetry technique was used to study density of complex eigenvalues of
``almost-Hermitian'' random matrices $X$ . These matrices were written in
the form 
\begin{equation}
\label{a11}X=A+i\alpha N^{-1/2}B 
\end{equation}
with $N\times N$ Hermitian statistically independent matrices $A$ and $B$,
and a number $\alpha $ of the order of unity. Due to the presence of the
factor $N^{-1/2}$ the ensemble described by Eq. (\ref{a11}) differs from
ensembles of matrices with arbitrary complex elements studied previously 
\cite{ginibre}. Fyodorov {\it et al} \cite{fyodorov} calculated the joint
probability of the real and imaginary parts of eigenvalues of the matrices $%
X $ corresponding to the the function $P\left( \epsilon ,y\right) $, Eq. (%
\ref{a3}).

Reducing in a standard way integration over the matrices $A$ and $B$ to
integration over supervectors and then to supermatrices, the authors of Ref. 
\cite{fyodorov} arrived in the limit $N\rightarrow \infty $ at the $0D$
version of the free energy $F\left[ Q\right] $, Eq. (\ref{a9}), with ${\bf h}%
^2\sim \alpha ^2$. The symmetry of $Q$ corresponded to the supermatrices $Q$
of Refs. \cite{efetov1,efetov} for the unitary ensemble, which is due to the
hermiticity of the matrices $A$ and $B$. Apparently, if they used real
symmetric matrices $A$, antisymmetric matrices $B$, and imaginary $\alpha $
they would obtain Eq. (\ref{a9}) with the supermatrices $Q$ corresponding to
the orthogonal ensemble (below these cases are called simply orthogonal and
unitary). This demonstrates that the models of disorder, Eqs. (\ref{a1},\ref
{a2}), in a limited volume are equivalent to the ensembles of weakly
non-symmetric (or non-Hermitian if the time reversal symmetry is broken)
random matrices.

For explicit calculations, Fyodorov {\it et al }\cite{fyodorov} used the
parametrization of Ref. \cite{efetov1}. However, due to the presence of the
new terms with ${\bf h}$ and $y$ in Eq. (\ref{a9}), the calculations with
this parametrization are very difficult. So, the final result was obtained
for the unitary ensemble only. As concerns the orthogonal case, computations
with this parametrization do not seem to be possible at all. At the same
time, study of the orthogonal ensemble can be very important because it
describes the vortices in superconductors and, as will be seen later, the
function $P\left( \epsilon ,y\right) $ for the orthogonal ensemble at small $%
h$ is qualitatively different from that for the unitary one.

Fortunately, one can circumvent the difficulties using a new
parametrization. Leaving details for another publication I want to present
here only a general structure. The supermatrix $Q$ is written as 
\begin{equation}
\label{a12}Q=ZQ_0\bar Z 
\end{equation}
with supermatrices $Z$ satisfying the conditions $Z\bar Z=1$ and $\left[
Z,\Lambda _1\right] =0$. The central part $Q_0$ is chosen as%
$$
Q_0=\left( 
\begin{array}{cc}
\cos \hat \phi & -\tau _3\sin 
\hat \phi \\ -\tau _3\sin \hat \phi & -\cos \hat \phi 
\end{array}
\right) ,\;\hat \phi =\left( 
\begin{array}{cc}
\phi & 0 \\ 
0 & i\chi 
\end{array}
\right) 
$$
The most lengthy part is calculation of the Jacobian but then the
computation is of no difficulties and the density of complex eigenenergies $%
P\left( \epsilon ,y\right) $ for the unitary ensemble takes the form 
\begin{equation}
\label{a13}P\left( \epsilon ,y\right) =\frac{\nu \sqrt{\pi }}{a\Delta }\exp
\left( -\frac{x^2}{4a^2}\right) \int_0^1\cosh xt\exp \left( -a^2t^2\right)
dt 
\end{equation}
where $x=2\pi y/\Delta $, $a^2=2\pi D_0h^2/\Delta $, $\Delta $ is the mean
level spacing.

Eq. (\ref{a13}) is exactly the result obtained in Ref. \cite{fyodorov}. For
the models of weak disorder, Eqs. (\ref{a1},\ref{a2}), $P\left( \epsilon
,y\right) $ depends only on the variable $y$ because $\nu $ is constant. For
the random matrix model, Eq. (\ref{a11}), $\nu $ depends on $\epsilon $ and
obeys the Wigner semi-circle law. In the limit $a\gg 1$ one obtains 
\begin{equation}
\label{a14}P\left( \epsilon ,y\right) \approx \frac{\pi \nu }{2a^2\Delta }%
,\;\left| x\right| \leq 2a^2 
\end{equation}
which means that the function $P\left( \epsilon ,y\right) $ is practically
constant within the interval $\left| x\right| \leq 2a^2$. Beyond this
interval the function exponentially decays. For the random matrix model this
behavior corresponds to the well known ``elliptic law'', Ref. \cite{ginibre}%
. For any $a$ the function $P\left( \epsilon ,y\right) $ is smooth, which
means, in particular, that the probability to find real eigenvalues is
negligible.

The situation drastically changes if the time-reversal symmetry is not
broken. Although now the computation of $P\left( \epsilon ,y\right) $ is
somewhat more lengthy, one can finally express this function in a rather
simple form 
$$
P\left( \epsilon ,y\right) =P_r+P_c,\;P_r=\nu \delta \left( y\right)
\int_0^1\exp \left( -a^2t^2\right) dt 
$$
\begin{equation}
\label{a15}P_c=\frac{\pi \nu }\Delta \Phi \left( \frac{\left| x\right| }{2a}%
\right) \int_0^1\exp \left( -a^2t^2\right) \sinh \left( \left| x\right|
t\right) tdt 
\end{equation}
where $\Phi \left( v\right) =\frac 2{\sqrt{\pi }}\int_v^\infty \exp \left(
-u^2\right) du$.

We see that the function $P\left( \epsilon ,y\right) $ contains the part $%
P_r $ proportional to $\delta \left( y\right) $. This means that a finite
fraction of eigenstates has real eigenenergies for any finite $a$. The
distribution of complex eigenenergies is described by the smooth function $%
P_c$. The fraction of the states with real eigenvalues vanishes only in the
limit $a\rightarrow \infty $. In this limit one comes for $\left| x\right|
\gg a$ to Eq. (\ref{a14}), which again results for the random matrix model
in the elliptic law.

The result about the finiteness of the fraction of eigenstates with real
eigenenergies as well as Eq.(\ref{a15}) itself is new although indications
of a peculiar behavior of the probability of real eigenvalues for asymmetric
real matrices can be found in Ref.\cite{ginibre}. In
Ref.\cite{hatano} a mixture of states with real and complex eigenvalues was
found numerically near the band center for the $2D$ model, Eq.(\ref
{a2}). Perhaps, the localization length in that region of
parameters exceeded the sample size. This would correspond to the $0D$
situation rather than $2D$. If it is so, the analytical result agrees with
the numerical observation.

Study of the directed disorder problems in higher dimensions, although 
more difficult, is definitely of interest. Solution of the $\sigma $-model,
Eqs. (\ref{a14},\ref{a15}), in one dimension seem to be possible using the
transfer-matrix method. This case corresponds to a thick film in a parallel
magnetic field in the problem of the vortices in a superconductor. It would
be interesting to obtain a localization-delocalization transition
analytically. The one-dimensional $\sigma $-model is also known to describe
the quantum kicked rotor problem \cite{fyodorov1}. May be, finite $h$
correspond to a dissipation in this problem and one can find a transition
here, too.

In summary, quantum disorder problems with a direction (imaginary
vector-potential) are considered. It is demonstrated that they can be mapped
onto a new supermatrix $\sigma $-model. The $0D$ version of the model
describes a class of phenomena that can be called directed quantum chaos.
Using the $0D$ $\sigma $-model the equivalence of this class of systems to
ensembles of weakly asymmetric or non-Hermitian random matrices is
established. The joint probability of complex eigenvalues is computed
explicitly and it is discovered that the fraction of real eigenenergies for
time reversal invariant systems is always finite.


\begin{references}
\bibitem{fogedby}  H.C. Fogedby, A.B. Eriksson, and L.V. Mikheev, Phys. Rev.
Lett. {\bf 75}, 1883 (1995)

\bibitem{verb}  J.J.M. Verbaarschot, H.A. Weidenm\"uller, and M.R.
Zirnbauer, Phys. Rep. {\bf 129}, 367 (1985)

\bibitem{somp}  H. Sompolinsky, A. Crisanti, and H.-J. Sommers, Phys. Rev.
Lett. {\bf 61}, 259 (1988)

\bibitem{hatano}  N. Hatano, D. Nelson, Phys. Rev. Lett. {\bf 77}, 570 (1996)

\bibitem{civale}  L. Civale {\it et al}, Phys. Rev. Lett. {\bf 67}, 648
(1991){\it \ }

\bibitem{nelson}  D.R. Nelson and V. Vinokur, Phys. Rev. B {\bf 48}, 13 060
(1993), and references therein.

\bibitem{zirn}  M.R. Zirnbauer, Nucl. Phys. A{\bf 560}, 95 (1993)

\bibitem{obukhov}  S.P. Obukhov, Physica A{\bf 101}, 145 (1980)

\bibitem{efetov}  K.B. Efetov, {\it Supersymmetry in Disorder and Chaos},
Cambridge University Press, New York (1997)

\bibitem{efetov1}  K.B. Efetov, Adv. in Phys. {\bf 32}, 53 (1983)

\bibitem{muz}  B.A. Muzykantskii and D.E. Khmelnitskii, JETP Lett. {\bf 62},
76 (1995); A.V. Andreev, O. Agam, B.D. Simons, and B.L. Altshuler, Phys.
Rev. Lett. {\bf 76}, 3947 (1996)

\bibitem{fyodorov}  Y.V. Fyodorov, B.A. Khoruzhenko, and H.-J. Sommers,
preprint, cond-mat/9606173

\bibitem{ginibre}  J.Ginibre, J.Math.Phys.{\bf 6,} 440; M.L.Mehta, {\it %
Random Matrices,} Academic Press, San Diego (1991); F.Haake, {\it Quantum
Signatures of Chaos,} Springer, Berlin (1991); N.Lehmann, H.-J.Sommers,
Phys.Rev.Let.{\bf 67}, 941 (1991)

\bibitem{fyodorov1}  Y.V. Fyodorov, A.D. Mirlin, Phys. Rev. Lett. {\bf 67},
2405 (1991); A. Altland, M. Zirnbauer, Phys. Rev. Lett. {\bf 77}, 4536 (1996)
\end{references}
\end{document}